\documentclass[prb,preprint,amsmath,amssymb]{revtex4}

\usepackage{graphicx}
\usepackage{dcolumn}
\usepackage{bm}

\begin{document}

\title{Macroscopic evidence of quantum coherent oscillations of the total spin in the
Mn-[3x3] molecular nanomagnet.}

\author{S. Carretta,$^1$ P. Santini,$^1$
E. Liviotti,$^1$ N. Magnani,$^1$ R. Caciuffo,$^2$ T. Guidi,$^2$
and G. Amoretti.$^1$ }
\affiliation{$^{1}$Istituto Nazionale per
la Fisica della Materia, Dipartimento di Fisica, Universit\`a di
Parma, I-43100 Parma, Italy} \affiliation{$^{2}$Istituto Nazionale
per la Fisica della Materia, Universit\`a Politecnica delle
Marche, I-60131 Ancona, Italy}
\date{October 2, 2003}
\begin{abstract}
Molecular nanomagnets, besides promising to open new frontiers in
technology, have attracted huge interest in the scientific
community because they can exhibit the phenomenon known as quantum
tunnelling of the magnetization, i.e. coherent fluctuations of the
direction of the total spin vector. In this paper we study a
different quantum phenomenon involving fluctuations of the
magnitude of the total spin vector. These fluctuations are related
to the mixing between states with different spin quantum number,
and imply new macroscopic effects, which we theoretically
investigated in the Mn-[3x3] grid.
\end{abstract}

\maketitle

Molecular nanomagnets (MNMs) \cite{Sessoli93,Gatteschi94,Legget95}
are clusters containing a finite number of transition-metal ions
whose magnetic moments (spins) are so strongly coupled that at low
temperature each molecule behaves like a single-domain particle
with fixed total spin. Being at the crossover between classical
and quantum regimes, MNMs exhibit at the same time classical
properties of macroscopic magnets such as magnetization
hysteresis, and quantum phenomena like tunneling of the direction
of the total spin through energy
barriers\cite{Friedman96,Thomas96,Wernsdorfer99}. MNM systems are
interesting also for potential technological applications, as
envisaged for the implementation of quantum computing algorithms
\cite{Leuenberger01}, or for dense and highly efficient memory
devices \cite{Sessoli93}. Here we study a new macroscopic
manifestation of a quantum phenomenon involving fluctuations not
only of the direction, but also of the magnitude of the total
spin, and we show that it is realized in a Mn-[3$\times$3] grid
\cite{Thompson00,Wald02}. Recognizing the effects of these
fluctuations is essential to achieve a satisfactory understanding
of the role played by quantum mechanics in complex mesoscopic
magnetic systems. Quantum magnetic phenomena were identified even
in molecules of great biological interest such as
ferritin\cite{ferritin1,ferritin2}.

The advantage of studying quantum phenomena in MNMs is that
the vanishingly small interaction between different molecules
allows single-molecule phenomena to be observed at a macroscopic
scale, because the crystal behaves like a collection of independent
objects \cite{Gatteschi94,Wald97}, each described in general by
spin Hamiltonians of the form \cite{Bencini90}

$$
H=\sum_{i>j}J_{ij}{\bf s}_{i}\cdot {\bf s}_{j} + \sum_{i}{\bf
s}_{i}\cdot {\bf D}_{i}\cdot {\bf s}_{i} +
$$
$$
\sum_{i>j}{\bf s}_{i}\cdot {\bf D}_{ij}\cdot {\bf s}_{j}+ \mu_B
\sum_{i}g_i{\bf B}\cdot {\bf s}_{i},\eqno(1)
$$

where ${\bf s}_{i}$ are spin operators of the $i^{th}$ ion in the
molecule. The first term is the isotropic Heisenberg exchange
interaction, the second and third terms describe the local
crystal-field and the anisotropic intra-cluster spin-spin
interactions. The last term is the Zeeman coupling with an
external field ${\bf B}$ in which isotropic $g$-factors are
assumed. While the Heisenberg term is rotationally invariant and
therefore conserves the length $\vert {\bf S}\vert$ of the total
spin ${\bf S} = \sum_{i}{\bf s}_{i}$, the anisotropic terms break
rotational invariance and do not conserve this observable.
Nevertheless, since the Heisenberg contribution is usually largely
dominant, $\vert {\bf S}\vert$ is nearly conserved, and the energy
spectrum of $H$ consists of a series of level multiplets with an
almost definite value of $\vert {\bf S}\vert$ (expressed in terms
of the quantum number $S$ as $\sqrt{S(S+1)}$). Thus, quantum
fluctuations of $\vert {\bf S}\vert$, which are associated with
mixing of states with different value of the quantum number $S$
(''$S$-mixing" \cite{Liviotti02}), either are zero or are expected
to produce negligible effects on the macroscopic behavior, and are
therefore neglected in virtually all studies. A major theoretical
goal would be to identify a clear macroscopic signature of such
fluctuations, and a model system displaying such effect.

In this paper we show that favorable conditions are met in the
Mn(II)-[3$\times$3] grid-like cluster
[Mn$_9$(2POAP-2H)$_6$](ClO$_4$)$_6$$\cdot$3.57MeCN$\cdot$H$_2$O
(hereafter Mn-[3$\times$3], see Fig. 1), a system that has been
recently characterized by magnetization and torque measurements
\cite{Wald02,torque}. Our theoretical calculations suggest
grid-shaped molecules as good candidates to study fluctuations of
$\vert {\bf S}\vert$, since the lowest level is expected to
display significant $S$-mixing with the first excited multiplet.
In fact, unlike in ideal ring-shaped molecules \cite{Waldepl}, in
the [3$\times$3] grid the two lowest manifolds of the exchange
part of the spin Hamiltonian belong to the same irreducible
representation of the molecular symmetry group. Therefore, the
application of a suitably oriented magnetic field induces a series
of anticrossings (ACs) between the ground state and levels
originating from higher excited manifolds (see red lines in Fig.
2). At the AC fields, $S$-mixing in the ground state is maximum,
and quantum fluctuations of $\vert {\bf S}\vert$ are greatly
enhanced. Torque\cite{Wald02} and neutron\cite{GuidiUn}
experiments on Mn-[3$\times$3] show that the zero-field gap
between the two lowest $S$-multiplets is small enough for the ACs
to occur at fields within experimental reach.

Mn-[3$\times$3] crystallizes in the space group $C_2/c$, and the
cation [Mn$_9$(2POAP-2H)$_6$]$^{6+}$ exhibits a slightly distorted
$S_4$ molecular symmetry with the $C_2$ axis perpendicular to the
plane of the cluster \cite{Thompson00}. The average distance
between the Mn(II) ions is 3.93 $\AA$, the smallest distance
between clusters is larger than 8 $\AA$. For a cluster composed of
nine interacting Mn(II) spins with ${\bf s}_{i} = 5/2$ the
dimension of the Hilbert space is 10077696. The difficulties
related with this huge dimension have been overcome by exploiting
both the irreducible tensor operator technique and the Lanczos
algorithm for the exact diagonalisation. The two-step procedure
already developed \cite{Carretta03} has allowed the inclusion of
$S$-mixing effects in the calculation. The exchange integrals and
the single-site and spin-spin anisotropy tensors have been
determined by neutron spectroscopy \cite{GuidiUn}. All exchange
integrals between nearest-neighbors are found to be nearly equal
to $0.47$ meV, apart from $J_{18}$, $J_{78}$, $J_{34}$ and
$J_{45}$ (see Fig. 1), whose values are $0.33$ meV.
Next-nearest-neighbor exchange interactions can be neglected.
Concerning the local crystal-field, the second term in Eq. 1 can
be approximately rewritten as

$$
\sum_{i}{{\bf s}_{i}}\cdot {{\bf D}_{i}}\cdot {{\bf s}_{i}}=D
\sum_{i}{[s_{iz}^2 - \frac{1}{3} s_i(s_i+1)]},\eqno(2)
$$

with $D=-6.1$ $\mu$eV. The intracluster dipole-dipole interaction
${\bf D}_{ij}$ has been evaluated within the point-dipole
approximation\cite{Bencini90}. At last, $g_i=2$ is assumed, as
appropriate for Mn(II) ions.

With these experimentally determined parameters, the energy
spectrum for ${\bf B}=0$ consists of many level multiplets with an
almost definite value of $S$. The ground multiplet has $S=5/2$,
and the four lowest-lying excited multiplets have (in order of
increasing energy) $S=7/2$, $S=3/2$, $S=3/2$, $S=9/2$. These
multiplets are separated by the isotropic exchange and split by
the anisotropic interactions. The lowest levels are shown in Fig.
2. The application of a magnetic field $\bf{B}$ in a direction
outside the grid plane and different from that of the $C_2$ axis
produces several ACs involving levels belonging to different
multiplets (see Fig. 2). As the AC fields $\bf{B}_c$ are
approached the multiplet mixing is enhanced. For
$\bf{B}=\bf{B}_c$, the spins in each cluster oscillate coherently
between states with different values of $S$, which therefore is no
longer a good quantum number. This can be inferred for example for
the ground state ACs (indicated by the arrows in Fig. 2) from the
inset in Fig. 2, which shows the field-dependence at $T=0$ of the
quantity $S_{eff}$, defined through the relation
$S_{eff}(S_{eff}+1) = \langle {\bf S}^2\rangle$. When $\bf{B}$ is
close to $\bf{B}_c$ the value of $S_{eff}$ is intermediate between
half integers, e.g. 5/2 and 7/2 at the first level AC, thus
confirming that the ground-state wavefunction is a superposition
of different total-spin states.

This opens a scenario in which not only the direction of the total
spin fluctuates in time, as in the case of the well-studied
quantum tunnelling of the magnetization
\cite{Friedman96,Thomas96,Wernsdorfer99}, but even its length
fluctuates (quantum dynamics of the total spin, or briefly QDTS).
Moreover, by properly tuning the direction of the applied magnetic
field, the AC splitting can be made as large as several Kelvins,
thus overcoming the problem of decoherence due to hyperfine fields
and to cluster-cluster interactions. Most importantly, in the
Mn-[3$\times$3] grid several of these ACs involve the two lowest
energy levels, thus opening the possibility to detect the QDTS by
means of ${\it macroscopic}$ low-temperature magnetic bulk
techniques. In particular, our calculations show that at low
temperature, in correspondence to each AC between the two lowest
lying states, a sharp peak should be detected in the torque signal
(as function of the field intensity $B$) when the direction of the
field $\theta$ is close to the grid plane ($\theta=0^o$) or almost
perpendicular to it ($\theta=90^o$) (see Figs. 3 and 4). The
precipitous drop of the torque signal as a function of $\theta$
near $\theta=0^o$ and $\theta=90^o$ is due to a change of sign
imposed by symmetry (as ${\bf M}\times {\bf B}=0$ when ${\bf B}$
points along a principal direction of the susceptibility tensor of
the system).

By comparing Figs. 3 and 4 with the corresponding spectrum
reported in Fig. 2, it is evident that a torque peak appears in
correspondence to each AC involving the ground state. While the
first peak at low field corresponds to an intra-multiplet AC, the
peaks at about $6.6$, $8.9$ and $11.2$ Tesla represent a direct
consequence of the coherent superposition of different total spin
quantum states. The link between the peaks in the torque signal
and the level ACs is demonstrated in Fig. 4, where the torque
curves calculated with the mixing artificially forced to zero are
shown by colored dashed lines. When $S$-mixing is neglected, every
inter-multiplet AC becomes a crossing, the fluctuations of the
total spin vector are suppressed and the peaks in the torque curve
disappear. Thus, without $S$-mixing the torque curve should
exhibit the step-like field dependence usually observed in MNMs
\cite{Waldtorque}. The comparison between solid and dashed curves
in Fig. 4 shows clearly that the predicted peaks in the torque
would be a direct macroscopic manifestation of the QDTS at the
level ACs. Also, these spin fluctuations and the associated peaks
are of purely quantum origin, as these are absent in the classical
version of the Hamiltonian Eq. (1) (see the gray dashed curve in
Fig. 4).

The fact that these torque peaks directly reflect the large
enhancement of quantum fluctuations of $\vert {\bf S}\vert$ at the
ACs can be understood by a simple physical picture. In the
following we define the $z'$ axis parallel to ${\bf B}$. The
torque is proportional to the magnetic response perpendicular to
the field direction, i.e. to $\langle S_{x'}\rangle$. Fluctuations
of $\vert {\bf S}\vert$ are accompanied by fluctuations of the
magnetization ($S_{z'}$)\cite{collinear}, and these latter are
deeply connected with $\langle S_{x'}\rangle$. Accordingly,
$S_{x'}$ tracks the increase and decrease of these fluctuations
while sweeping over the AC, leading to a peak in the torque.
Indeed, near the ACs, the ground-state wavefunction can be written
to a good degree of approximation as

$$
\vert G\rangle = \frac{1}{\sqrt{2}}(a(B)\vert \Gamma_1 ,M\rangle +
b(B)\vert \Gamma_2 ,M+1 \rangle ),\eqno(3)
$$

where $a(B)^2 + b(B)^2 = 2$ by normalization. For $B\ll B_c$,
$b(B)\sim 0$, $a(B)\sim\sqrt{2}$, and the ground state reduces
approximately to an eigenstate of $S_{z'}$, $\vert \Gamma_1
,M\rangle$, where $M$ is the corresponding eigenvalue and
$\Gamma_1$ represents the set of additional labels necessary to
identify the state. For $B\gg B_c$, $a(B)\sim 0$,
$b(B)\sim\sqrt{2}$, and the ground state reduces approximately to
$\vert \Gamma_2,M+1\rangle$.

At $T=0$ the torque is given by

$$
\tau \propto B\langle S_{x'}\rangle = B a(B) b(B) \langle
\Gamma_1, M \vert S_{x'}\vert \Gamma_2, M+1\rangle.\eqno(4)
$$
Since $\langle \Gamma_1, M \vert S_{x'}\vert \Gamma_2, M+1\rangle$
is almost independent on $B$, the field dependence of $\tau$ comes
entirely from the factor $B a(B) b(B) = B b(B)\sqrt{2-b(B)^2}$.

Quantum fluctuations of the magnetization are usually
characterized by the quantity

$$
(\Delta S_{z'})^2 = \langle S_{z'}^2\rangle - \langle
S_{z'}\rangle^2 = \frac{b(B)^2}{2}-\frac{b(B)^4}{4}.\eqno(5)
$$
Therefore,
$$
\tau \propto 2 B \Delta S_{z'}.\eqno(6)
$$

In case of anticrossing, $\Delta S_{z'}$ is maximum at $B_c$,
where $b(B_c) = a(B_c) = 1$, and the torque peaks. In case of
crossing $\Delta S_{z'}$ is always zero because either $b(B)=0$ or
$a(B)=0$, and the torque does not peak. As a further confirmation
of this picture we show in Fig. 5 numerical calculations of the
fluctuations of $S_{z'}$ (''parallel") and of $S_{x'}$ (''
perpendicular"), with and without $S-mixing$.

The torque is the macroscopic magnetic bulk technique displaying
the clearest signature of $S$-mixing. When the temperature $T$
increases these effects wash out, and indeed published torque
measurements on Mn-[3$\times$3] \cite{Wald02} do not display peaks
because data were collected at too high a temperature.

We have also calculated the low-$T$ heat capacity as a function of
the applied field (see Fig. 6). This quantity provides a way to
assess the value of the AC gaps as a function of the field
direction. For $\theta \ne 0$ anticrossings open up and the heat
capacity does not vanish at $B_c$.

In this paper we have identified a new clear-cut macroscopic
manifestation of a quantum phenomenon characterizing the
microscopic dynamics of magnetic clusters. We have shown that
torque measurements provide direct evidence of quantum
fluctuations of the total spin length, and we have studied a model
system displaying such effect. The comprehensive understanding of
the spin dynamics in mesoscopic systems like molecular nanomagnets
is essential to understand the crossover between quantum and
classical mechanics and to make rational design of these compounds
possible, especially in view of the promising applications.

\newpage
{\bf Fig. 1}: A schematic representation of the molecular
structure of the Mn-[3$\times$3] grid showing the network of
manganese (blue) and oxygen (green) bonds. Other atoms are omitted
for clarity. The red arrows indicate the spin directions in the
classical ground state.

\vspace{2cm}

{\bf Fig. 2}: Calculated energy levels of the Hamiltonian Eq.(1)
with the parameters given in the text, and with the direction of
the applied field forming an angle $\theta = 2.8^o$ with the grid
plane. Energies are plotted as functions of the applied field
intensity $B$. The ground state energy is set equal to zero.
Arrows indicate the anticrossings produced by $S$-mixing between
the two lowest levels (highlighted in red). The inset shows
$S_{eff}$ as functions of $B$ for the same angle at $T=0$, where
$S_{eff}$ is defined through the relation $S_{eff}(S_{eff}+1) =
\langle {\bf S}^2\rangle$.

\vspace{2cm}

{\bf Fig. 3}: Two-dimensional plot of the calculated torque as a
function of the applied field direction $\theta$ and intensity
$B$ at $T=0.05$ K with parameters of Hamiltonian Eq.(1) as given
in the text.

\vspace{2cm}

{\bf Fig. 4}: Plots of the calculated torque at $T=0.4$ K as a
function of the applied magnetic field $B$ for two selected
directions $\theta$ lying close to the grid plane. Colored dashed
lines represent the calculated torque at the same angles with
$S$-mixing eliminated by hand, i.e. with quantum fluctuations of
$\vert {\bf S} \vert$ removed. The dashed gray line represents the
torque calculated using the classical version of the Hamiltonian
Eq. (1) at $T=0$ for $\theta = 2.8^o$. The classical value of the
torque has been rescaled (by about 3) to fit the figure. The inset
shows the calculated torque at $T=0.4$ K as a function of the
applied field intensity $B$ for several directions $\theta$.

\vspace{2cm}

{\bf Fig. 5}: Numerical calculations of the $T=0$ fluctuations of
$S_{z'}$ (''parallel") and of $S_{x'}$ ('' perpendicular"), as a
function of the applied magnetic field (forming an angle $\theta =
2.8^o$ with the grid plane), with and without $S-mixing$.
Parameters are given in the text.

\vspace{2cm}

{\bf Fig. 6}: Numerical calculations of the heat capacity as a
function of the applied magnetic field for several directions
$\theta$, at $T=0.4$ K.


\begin{references}
\bibitem{torque} The torque ${\bf \tau}$ experienced by a
magnetically anisotropic material in a uniform magnetic field
${\bf B}$ is given by ${\bf \tau}={\bf M}\times {\bf B}$, where
${\bf M}$ is the magnetization of the sample.
\bibitem{collinear} This should
be more properly named collinear magnetization.
\end{references}
\end{document}